# Exact Safety Verification of Hybrid Systems Based on Bilinear SOS Representation*

Zhengfeng Yang[a], Min Wu[a] and Wang Lin[a,b]

[a] Shanghai Key Laboratory of Trustworthy Computing
East China Normal University, Shanghai 200062, China
[b] College of Mathematics and Information Science
Wenzhou University, Zhejiang 325035, China
{zfyang,mwu}@sei.ecnu.edu.cn; linwang@wzu.edu.cn

**Abstract**

In this paper, we address the problem of safety verification of nonlinear hybrid systems. A hybrid symbolic-numeric method is presented to compute exact inequality invariants of hybrid systems efficiently. Some numerical invariants of a hybrid system can be obtained by solving a bilinear SOS programming via PENBMI solver or iterative method, then the modified Newton refinement and rational vector recovery techniques are applied to obtain exact polynomial invariants with rational coefficients, which *exactly* satisfy the conditions of invariants. Experiments on some benchmarks are given to illustrate the efficiency of our algorithm.

## 1. Introduction

Complex physical systems are systems in which the techniques of sensing, control, communication and coordination are involved and interacted with each other. Since many of such systems are safety critical systems, such as the controllers widely used in airplanes, railway, and automobiles, ensuring correct functioning of these systems is among the most important and challenging problems in computer science, mathematics and engineering. As a common mathematical model for complex physical systems, hybrid systems [6] are dynamical systems that are governed by interacting discrete and continuous dynamics. Continuous dynamics is specified by differential equations, and for discrete transitions, the hybrid system changes state instantaneously and possibly discontinuously. Among the most important research issues on hybrid systems are those of *safety*, i.e., deciding whether a given property holds in all the reachable states, and its dual problem *reachability*, i.e., deciding if there exists a trajectory starting from the initial set that reaches a state satisfying the given property.

Due to the infinite number of possible states in continuous state spaces, safety verification or reachability analysis of hybrid systems presents a challenge. Some well-established techniques have been proposed. In [2], level set methods and flow-pipe approximations were presented for computing *approximate* reachable sets of hybrid systems. By contrast, quantifier elimination was used in [11] to compute *exact* reachable sets for linear systems with certain eigenstructures and semi-algebraic initial sets, and this method was generalized in [28] to handle linear systems with almost arbitrary eigenstructures. Recently, invariant generation has been proposed for safety verification of hybrid systems. An *invariant* [25] of a hybrid system is a property that holds in all the reachable states of the system, in other words, it is an over-approximation of all the

*This material is supported in part by the Chinese National Natural Science Foundation under Grants 91018012,61021004(Yang, Wu) and 10801052(Wu) and 10901055(Yang), and the Scientific Research Project of The Graduate School of East China Normal University under Grant CX2011009.

reachable states. Invariants are useful facts about the dynamics of a given system. However, generating invariants of arbitrary form is known to be computationally hard, intractable even for the simplest classes. The usual technique for generating invariants is to compute an *inductive invariant*, an assertion that holds initially and is preserved by all discrete and continuous state changes. There has been lots of work towards invariant generation for hybrid systems using convex optimization and semi-algebraic system solving [24, 5, 16, 19, 25, 20, 28, 27]. However, some of these techniques are subject to numerical errors and some suffer from high complexity. Taking advantage of the efficiency of numerical computation and the error-free property of symbolic computation, we proposed in [29] a hybrid symbolic-numeric method via sum of squares (SOS) relaxation and exact certificate to construct differential invariants for continuous dynamic systems, and generalized in [15] the idea for safety verification of hybrid systems.

In this paper, we present a hybrid symbolic-numeric algorithm to compute exact invariants of hybrid systems. The algorithm is based on SOS relaxation [17] of a parametric polynomial optimization problem with bilinear matrix inequality (BMI) constraints, which can be solved directly using a recently developed PENBMI solver [9, 10] in Matlab or an iterative method, and exact SOS representation recovery techniques presented in [7, 8]. The algorithm in this work improves our former result in [15] in two aspects. First, we replace the strengthened linear matrix inequality (LMI) constraints with the original BMI constraints in the parametric optimization problem. Second, we modify both Newton iteration refinement and rational vector recovery, which can handle some cases where the method in [15] fails and usually yield invariants with lower degree. Unlike the numerical approaches, our method can yield exact invariants, which can overcome the unsoundness in verification of hybrid systems caused by numerical errors [18], as illustrated in Example 2. In comparison with some symbolic approaches based on qualifier elimination technique, our approach is more efficient and practical, because parametric polynomial optimization problem based on SOS relaxation can be solved in polynomial time theoretically.

The rest of the paper is organized as follows. In Sections 2, we introduce some notions about hybrid systems and safety verification. And in Section 3, we transform the problem of safety verification of hybrid systems into a parametric program with BMI constraints. Section 4 is devoted to computing numerical solutions of a BMI problem via PENBMI solver or iterative method. In Section 5, an algorithm based on the modified Newton iteration and rational vector recovery techniques is proposed to obtain exact solutions of BMI problem with rational coefficients. In Section 6, we discuss issues related to the implementation of the proposed mehtod. In Section 7, experiments on some benchmarks are shown to illustrate our algorithm on safety verification. Section 8 concludes the paper.

## 2. Hybrid Systems and Safety Verification

To model hybrid systems, we use the notion of hybrid automata [6, 25].

**Definition 1 (Hybrid system).** *A hybrid system* $\mathbf{H} : \langle V, L, \mathcal{T}, \Theta, \mathcal{D}, \Psi, \ell_0 \rangle$ *consists of the following components:*

- $V = \{x_1, ..., x_n\}$, *a set of real-valued system* variables. *A* state *is an interpretation of* $V$, *assigning to each* $x_i \in V$ *a real value. An* assertion *is a first-order formula over* $V$. *A state $s$ satisfies an assertion* $\varphi$, *written as* $s \models \varphi$, *if* $\varphi$ *holds on the state $s$. We will also write* $\varphi_1 \models \varphi_2$ *for two assertions* $\varphi_1, \varphi_2$ *to denote that* $\varphi_2$ *is true at least in all the states in which* $\varphi_1$ *is true;*

- $L$, *a finite set of locations;*

- $\mathcal{T}$, *a set of (discrete) transitions. Each transition* $\tau : \langle \ell, \ell', g_\tau, \rho_\tau \rangle \in \mathcal{T}$ *consists of a prelocation* $\ell \in L$, *a postlocation* $\ell' \in L$, *the guard condition* $g_\tau$ *over* $V$, *and an assertion* $\rho_\tau$ *over* $V \cup V'$ *representing the next-state relation, where* $V' = \{x'_1, ..., x'_n\}$ *denotes the nextstate variables. Note that the transition* $\tau$ *can take place only if* $g_\tau$ *holds;*

- $\Theta$, *an assertion specifying the* initial *condition;*

- $\mathcal{D}$, *a map that associates each location* $\ell \in L$ *to a* differential rule *(also known as a* vector field*)* $\mathcal{D}(\ell)$, *an autonomous system* $\dot{x}_i = f_{\ell,i}(V)$ *for each* $x_i \in V$, *written briefly as* $\dot{\mathbf{x}} = \mathbf{f}_\ell(\mathbf{x})$. *The differential rule at a location specifies how the system variables evolve in that location;*

- $\Psi$, *a map that maps each location* $\ell \in L$ *to a* location condition (location invariant) $\Psi(\ell)$, *an assertion over* $V$;

- $\ell_0 \in L$, *the* initial location. *We assume that the initial condition satisfies the location invariant at the initial location, that is,* $\Theta \models \Psi(\ell_0)$.

Given a hybrid system **H** with a prespecified unsafe region $X_u \subset \mathbb{R}^n$, we say that the system **H** is *safe* if all trajectories of **H** starting from any state in the initial set, can not evolve to $X_u$, or, equivalently, any state in $X_u$ is not reachable. We can also specify an unsafe region, denoted as $X_u(\ell)$, for each location $\ell \in L$.

For safety verification of hybrid systems, the notion of invariants of hybrid systems plays an important role. We define

**Definition 2 (Invariant).** *[25] An* invariant *of a hybrid system at location* $\ell$ *is an assertion* $\mathcal{I}$ *such that for any reachable state* $\langle \ell, \mathbf{x} \rangle$ *of the hybrid system,* $\mathbf{x} \models \mathcal{I}$.

*An* invariant *of a hybrid system is an assertion that holds in all the reachable states of the system.*

Clearly, an invariant of a hybrid system is an over-approximation of all the reachable states of the system. If an invariant lies inside the safe regions, or its intersection with the unsafe regions is empty, then safety of hybrid systems is verified. However, generating invariants with arbitrary form is known to be computationally hard, intractable even for the simplest classes. The usual technique for generating invariants is to compute inductive invariants, as defined below.

**Definition 3 (Inductive invariant).** *An* inductive assertion map $\mathcal{I}$ *of a hybrid system* **H** : $\langle V, L, \mathcal{T}, \Theta, \mathcal{D}, \Psi, \ell_0 \rangle$ *is a map that associates with each location* $\ell \in L$ *an assertion* $\mathcal{I}(\ell)$ *that holds initially and is preserved by all discrete transitions and continuous flows of* **H**. *More formally, an inductive assertion map satisfies the following requirements:*

**[Initial]** $\Theta \models \mathcal{I}(\ell_0)$.

**[Discrete Consecution]** *For each discrete transition* $\tau : \langle \ell, \ell', g_\tau, \rho_\tau \rangle$ *starting from a state satisfying* $\mathcal{I}(\ell)$, *taking* $\tau$ *leads to a state satisfying* $\mathcal{I}(\ell')$, *i.e.,* $\mathcal{I}(\ell) \wedge g_\tau \wedge \rho_\tau \models \mathcal{I}(\ell')$ *where* $\mathcal{I}(\ell')$ *represents the assertion* $\mathcal{I}(\ell)$ *with the current state variables* $x_1, \ldots, x_n$ *replaced by the next state variables* $x'_1, \ldots, x'_n$, *respectively.*

**[Continuous Consecution]** *For every location* $\ell \in L$ *and states* $\langle \ell, \mathbf{x}_1 \rangle$, $\langle \ell, \mathbf{x}_2 \rangle$ *such that* $\mathbf{x}_2$ *evolves from* $\mathbf{x}_1$ *according to the differential rule* $\mathcal{D}(\ell)$ *at* $\ell$, *if* $\mathbf{x}_1 \models \mathcal{I}(\ell)$ *then* $\mathbf{x}_2 \models \mathcal{I}(\ell)$.

By a *polynomial hybrid system*, we mean a hybrid system **H** : $\langle V, L, \mathcal{T}, \Theta, \mathcal{D}, \Psi, \ell_0 \rangle$, where the initial condition $\Theta$, location invariants $\Psi(\ell)$, and the guard condition and reset relation in each transition $\tau \in \mathcal{T}$ are conjunctions of polynomial inequalities over the program variables, and moreover, each differential rule $\mathcal{D}(\ell)$ is of the form $\dot{x}_i = f_{\ell,i}(\mathbf{x})$ with $f_{\ell,i}(\mathbf{x}) \in \mathbb{R}[\mathbf{x}]$.

In a preceding paper [15], we propose a symbolic-numeric approach to generate polynomial invariants of the form $\varphi(\mathbf{x}) \geq 0$ for polynomial hybrid systems via the combination of Sum-of-Squares (SOS) relaxation with Gauss-Newton refinement and rational vector recovery techniques. We will describe how to improve our result in [15] by solving the BMI problem directly. As stated in the following theorem, safety verification of hybrid systems can be reduced to finding invariants of hybrid systems.

**Theorem 1.** *[Theorem 2.7, [20]] Let* **H** : $\langle V, L, \mathcal{T}, \Theta, \mathcal{D}, \Psi, \ell_0 \rangle$ *be a hybrid system. Suppose for each location* $\ell \in L$, *there exists a function* $\varphi_\ell(\mathbf{x})$ *such that the following conditions hold:*

(i) $\Theta \models \varphi_{\ell_0}(\mathbf{x}) \geq 0$,

(ii) $\varphi_\ell(\mathbf{x}) \geq 0 \wedge g(\ell, \ell') \wedge \rho(\ell, \ell') \models \varphi_{\ell'}(\mathbf{x}') \geq 0$, for any transition $\langle \ell, \ell', g, \rho \rangle$ going out of $\ell$,

(iii) $\varphi_\ell(\mathbf{x}) = 0 \wedge \Psi(\ell) \models \dot{\varphi}_\ell(\mathbf{x}) > 0$, here $\dot{\varphi}_\ell(\mathbf{x})$ denotes the Lie-derivative of $\varphi_\ell(\mathbf{x})$ along the vector field $\mathcal{D}(\ell)$, i.e., $\dot{\varphi}_\ell(\mathbf{x}) = \sum_{i=1}^n \frac{\partial \varphi_\ell}{\partial x_i} \cdot f_{\ell,i}(\mathbf{x})$, then $\varphi_\ell(\mathbf{x}) \geq 0$ is an invariant of the hybrid system $\mathbf{H}$ at location $\ell$. If, moreover,

(iv) $X_u(\ell) \models \varphi_\ell(\mathbf{x}) < 0, \quad \forall \ell \in L$,

then the safety of the system $\mathbf{H}$ is guaranteed.

*Proof.* The proof is obvious. □

Clearly, to guarantee safety of the hybrid system $\mathbf{H}$, the intersection of the initial set $\Theta$ and the unsafe region $X_u(\ell_0)$ must be empty. When the hybrid system $\mathbf{H}$ in Theorem 1 specializes to a continuous system at one location, denoted as $\mathbf{D} : \langle V, \Theta, \mathcal{D}, \Psi \rangle$, then the condition (iv) in Theorem 1 can be relaxed through replacing the whole unsafe region $X_u$ by its boundary $\partial X_u$, as illustrated by the following.

**Corollary 1.** *Let $\mathbf{D} : \langle V, \Theta, \mathcal{D}, \Psi \rangle$ be a continuous system. Suppose there exists a function $\varphi(\mathbf{x})$ satisfying the following conditions:*

(i) $\Theta \models \varphi(\mathbf{x}) \geq 0$,

(ii) $\varphi(\mathbf{x}) = 0 \wedge \Psi \models \dot{\varphi}(\mathbf{x}) > 0$,

(iii) $\partial X_u \models \varphi(\mathbf{x}) < 0$, here $\partial X_u$ denotes the boundary of the set $X_u$,

*then the safety of the system $\mathbf{D}$ is guaranteed.*

*Proof.* By conditions (i) and (ii), the values of $\varphi(\mathbf{x})$ can not be negative during the entire evolution of the system $\mathbf{D}$. Then condition (iii) implies that all reachable sets lie outside the unsafe region $X_u$, yielding the safety of the system. □

In addition, we can consider more complicated forms of invariants for safety verification, i.e., invariants that are conjunctions of several polynomial inequalities: $\bigwedge_i \varphi_{\ell,i}(\mathbf{x}) \geq 0$. For simplicity, we consider the invariants of $\mathbf{H}$ of the form:

$$\varphi_{\ell,1}(\mathbf{x}) \geq 0 \wedge \varphi_{\ell,2}(\mathbf{x}) \geq 0.$$

The following theorem provides a method to determine the invariants of the above form.

**Theorem 2.** *Let $\mathbf{H} : \langle V, L, \mathcal{T}, \Theta, \mathcal{D}, \Psi, \ell_0 \rangle$ be a hybrid system. Suppose for each location $\ell \in L$, there exist two functions $\varphi_{\ell,1}(\mathbf{x}), \varphi_{\ell,2}(\mathbf{x})$ satisfying the following conditions:*

(i) $\Theta \models \varphi_{\ell_0,1}(\mathbf{x}) \geq 0 \wedge \varphi_{\ell_0,2}(\mathbf{x}) \geq 0$,

(ii) $\varphi_{\ell,1}(\mathbf{x}) \geq 0 \wedge \varphi_{\ell,2}(\mathbf{x}) \geq 0 \wedge g(\ell, \ell') \wedge \rho(\ell, \ell') \models \varphi_{\ell',1}(\mathbf{x}') \geq 0 \wedge \varphi_{\ell',2}(\mathbf{x}') \geq 0$, for any transition $\langle \ell, \ell', g, \rho \rangle$ going out of $\ell$,

(iii) $\varphi_{\ell,1}(\mathbf{x}) = 0 \wedge \varphi_{\ell,2}(\mathbf{x}) \geq 0 \wedge \Psi(\ell) \models \dot{\varphi}_{\ell,1}(\mathbf{x}) > 0$,

(iv) $\varphi_{\ell,1}(\mathbf{x}) \geq 0 \wedge \varphi_{\ell,2}(\mathbf{x}) = 0 \wedge \Psi(\ell) \models \dot{\varphi}_{\ell,2}(\mathbf{x}) > 0$,

*then $\varphi_{\ell,1}(\mathbf{x}) \geq 0 \wedge \varphi_{\ell,2}(\mathbf{x}) \geq 0$ is an invariant of the hybrid system $\mathbf{H}$ at location $\ell$.*

*Proof.* It is easy to prove that Conditions (i), (ii) and (iii) imply that $\varphi_{\ell,1}(\mathbf{x}) \geq 0 \wedge \varphi_{\ell,2}(\mathbf{x}) \geq 0$ satisfies three requirements in Definition 3. □

Similarly, the following corollary shows that the invariant in Theorem 2 can guarantee the safety of hybrid systems.

**Corollary 2.** *Let $\mathbf{H} : \langle V, L, \mathcal{T}, \Theta, \mathcal{D}, \Psi, \ell_0 \rangle$ be a hybrid system, and $X_u(\ell) = \{\mathbf{x} \in \mathbb{R}^n : \zeta_\ell(\mathbf{x}) \geq 0\}$ denotes the unsafe region at location $\ell$. Suppose for each location $\ell \in L$, there exist two functions $\varphi_{\ell,1}(\mathbf{x})$, $\varphi_{\ell,2}(\mathbf{x})$ that satisfy the conditions (i-iii) in Theorem 2, and moreover,*

**(v)** $\varphi_{\ell,1}(\mathbf{x}) \geq 0 \wedge \varphi_{\ell,2}(\mathbf{x}) \geq 0 \models \zeta_\ell(\mathbf{x}) < 0$,

*then the safety of the system $\mathbf{H}$ is guaranteed.*

**Remark 1.** $\varphi_{\ell,1}(\mathbf{x}) \geq 0 \wedge \varphi_{\ell,2}(\mathbf{x}) \geq 0$ in Theorem 2 is an invariant of the hybrid system $\mathbf{H}$ at location $\ell$, but it is not ensured that either $\varphi_{\ell,1}(\mathbf{x}) \geq 0$ or $\varphi_{\ell,2}(\mathbf{x}) \geq 0$ is an invariant of $\mathbf{H}$ at the location $\ell$. This kind of invariants is helpful to look for invariants of the form $c_1 \leq f(\mathbf{x}) \leq c_2$ where $f(\mathbf{x})$ is a polynomial over $\mathbb{R}$.

## 3. Problem Reformulation

For brevity, we will abuse the notation $\varphi_\ell(\mathbf{x})$ to represent both the function $\varphi_\ell(\mathbf{x})$ and the invariant $\varphi_\ell(\mathbf{x}) \geq 0$. Clearly, when the functions $\varphi_\ell(\mathbf{x})$ at all locations are identical to $\varphi(\mathbf{x})$, then $\varphi(\mathbf{x})$ becomes an inductive invariant of the hybrid system. Therefore, in the sequel we only discuss how to find invariants $\varphi_\ell(\mathbf{x})$ for each location $\ell \in L$, and the problem of computing an inductive invariant $\varphi(\mathbf{x})$ can be handled similarly. Remark that the invariants $\varphi_\ell(\mathbf{x})$ or $\varphi(\mathbf{x})$ are also known as barrier certificates in [20].

Let us predetermine a template of polynomial invariants with the given degree $d$. We assume that $\varphi_\ell(\mathbf{x}) = \sum_\alpha c_\alpha \mathbf{x}^\alpha$, where $\mathbf{x}^\alpha = x_1^{\alpha_1} \cdots x_n^{\alpha_n}$, $\alpha = (\alpha_1, \ldots, \alpha_n) \in \mathbb{Z}_{\geq 0}^n$ with $\sum_{i=1}^n \alpha_i \leq d$, and $c_\alpha \in \mathbb{R}$ are parameters. We can rewrite $\varphi_\ell(\mathbf{x}) = \mathbf{c}_\ell^T \cdot T_\ell(\mathbf{x})$, where $T_\ell(\mathbf{x})$ is the (column) vector of all terms in $x_1, \ldots, x_n$ with total degree $\leq d$, and $\mathbf{c}_\ell \in \mathbb{R}^\nu$, with $\nu = \binom{n+d}{n}$, is the coefficient vector of $\varphi_\ell(\mathbf{x})$. In the sequel, we write $\varphi_\ell(\mathbf{x})$ as $\varphi_\ell(\mathbf{x}, \mathbf{c}_\ell)$ for clarity.

From Theorem 1, to verify the safety of hybrid system $\mathbf{H}$ it suffices to find the invariants $\varphi_\ell(\mathbf{x})$ at each location $\ell \in L$. The latter problem can be translated into the following problem

$$\left. \begin{array}{l} \text{find } \mathbf{c}_\ell \in \mathbb{R}^\nu, \quad \forall \ell \in L \\ \text{s.t.} \quad \Theta \models \varphi_{\ell_0}(\mathbf{x}, \mathbf{c}_{\ell_0}) \geq 0 \\ \qquad \varphi_\ell(\mathbf{x}, \mathbf{c}_\ell) \geq 0 \wedge g(\ell, \ell') \wedge \rho(\ell, \ell') \models \varphi_{\ell'}(\mathbf{x}', \mathbf{c}_{\ell'}) \geq 0 \\ \qquad \varphi_\ell(\mathbf{x}, \mathbf{c}_\ell) = 0 \wedge \Psi(\ell) \models \dot{\varphi}_\ell(\mathbf{x}, \mathbf{c}_\ell) > 0 \\ \qquad X_u(\ell) \models \varphi_\ell(\mathbf{x}, \mathbf{c}_\ell) < 0 \end{array} \right\} \quad (1)$$

Suppose that

$$\begin{cases} \Theta = \{\mathbf{x} \in \mathbb{R}^n : \bigwedge_{l=1}^q \theta_l(\mathbf{x}) \geq 0\}, \quad X_u(\ell) = \{\mathbf{x} \in \mathbb{R}^n : \bigwedge_{j=1}^p \zeta_{\ell,j}(\mathbf{x}) \geq 0\}, \\ \Psi(\ell) = \{\mathbf{x} \in \mathbb{R}^n : \bigwedge_{k=1}^r \psi_{\ell,k}(\mathbf{x}) \geq 0\}, \quad g(\ell, \ell') = \{\mathbf{x} \in \mathbb{R}^n : \bigwedge_{i=1}^s g_{\ell\ell',i}(\mathbf{x}) \geq 0\}, \\ \rho(\ell, \ell')(\mathbf{x}, \mathbf{x}') = \{\mathbf{x}' \in \mathbb{R}^n : \bigwedge_{u=1}^t \rho_{\ell\ell',u}(\mathbf{x}, \mathbf{x}') \geq 0\}, \end{cases}$$

where $\ell, \ell' \in L$, and $\theta_l(\mathbf{x})$, $\zeta_{\ell,j}(\mathbf{x})$, $\psi_{\ell,k}(\mathbf{x})$, $g_{\ell\ell',i}(\mathbf{x})$ and $\rho_{\ell\ell',u}(\mathbf{x}, \mathbf{x}')$ are polynomials over $\mathbb{R}$.

Let $f_i(\mathbf{x}) \in \mathbb{R}[\mathbf{x}]$ for $1 \leq i \leq m$ and $g(\mathbf{x}) \in \mathbb{R}[\mathbf{x}]$. Suppose that the set $\{\mathbf{x} \in \mathbb{R}^n : \bigwedge_{i=1}^m f_i(\mathbf{x}) \geq 0\}$ is compact. According to Stengle's Positivstellensatz, Schmüdgen's Positivstellensatz or Putinar's Positivstellensatz, if there exist SOS polynomials $\sigma_i \in \mathbb{R}[\mathbf{x}]$ for $i = 0, \ldots, m$, such that $g(\mathbf{x})$ can be written as $g(\mathbf{x}) = \sigma_0(\mathbf{x}) + \sum_{i=1}^m \sigma_i(\mathbf{x}) f_i(\mathbf{x})$, then the assertion

$$\bigwedge_{i=1}^m (f_i(\mathbf{x}) \geq 0) \models g(\mathbf{x}) > 0$$

holds. Therefore, the existence of SOS representations provides a sufficient and necessary condition of the strict positiveness of $g(\mathbf{x})$ on the set $\{\mathbf{x} \in \mathbb{R}^n : \bigwedge_{i=1}^m f_i(\mathbf{x}) \geq 0\}$. Moreover, the degree bound of those unknown SOS polynomials $\sigma_i$ is exponential in $n$, $\deg(g)$ and $\deg(f_i)$. For more details the reader can refer to [13]. Based on the above observation, the problem (1) can be transformed into an equivalent SOS programming of the form

$$
\begin{aligned}
\text{find} \quad & \mathbf{c}_\ell \in \mathbb{R}^\nu, \quad \forall \ell \in L \\
\text{s.t.} \quad & \varphi_{\ell_0}(\mathbf{x}, \mathbf{c}_{\ell_0}) = \sigma_0(\mathbf{x}) + \sum_{l=1}^{q} \sigma_l(\mathbf{x})\theta_l(\mathbf{x}), \\
& \varphi_{\ell'}(\mathbf{x}', \mathbf{c}_{\ell'}) = \lambda_{\ell\ell',0}(\mathbf{x}) + \sum_{i=1}^{s} \lambda_{\ell\ell',i}(\mathbf{x})g_{\ell\ell',i}(\mathbf{x}) \\
& \qquad + \sum_{u=1}^{t} \gamma_{\ell\ell',u}(\mathbf{x})\rho_{\ell\ell',u}(\mathbf{x}, \mathbf{x}') + \eta_{\ell\ell'}(\mathbf{x})\varphi_\ell(\mathbf{x}, \mathbf{c}_\ell), \\
& \dot{\varphi}_\ell(\mathbf{x}, \mathbf{c}_\ell) = \phi_{\ell,0}(\mathbf{x}) + \sum_{k=1}^{r} \phi_{\ell,k}(\mathbf{x})\psi_{\ell,k}(\mathbf{x}) + \nu_\ell(\mathbf{x})\varphi_\ell(\mathbf{x}, \mathbf{c}_\ell) + \epsilon_{\ell,1} \\
& -\varphi_\ell(\mathbf{x}, \mathbf{c}_\ell) = \mu_{\ell,0}(\mathbf{x}) + \sum_{j=1}^{p} \mu_{\ell,j}(\mathbf{x})\zeta_{\ell,j}(\mathbf{x}) + \epsilon_{\ell,2},
\end{aligned}
\qquad (2)
$$

where $\sigma_l(\mathbf{x}), \lambda_{\ell\ell',i}(\mathbf{x}), \gamma_{\ell\ell',u}(\mathbf{x}), \eta_{\ell\ell'}(\mathbf{x}), \phi_{\ell,k}(\mathbf{x}), \mu_{\ell,j}(\mathbf{x})$ are SOSes in $\mathbb{R}[\mathbf{x}]$, $\nu_\ell(\mathbf{x}) \in \mathbb{R}[\mathbf{x}]$, and $\epsilon_{\ell,1}, \epsilon_{\ell,2} \in \mathbb{R}_+$. In practice, to avoid the high computational complexity, we simply set up a truncated SOS programming by fixing a *priori* (much smaller) degree bound $2e$, with $e \in \mathbb{Z}_+$, of all the unknown polynomials. Consequently, the existence of a solution $\mathbf{c}_\ell$ of (2) can guarantee the safety property of the given system.

In the SOS programming (2), the decision variables are the coefficients of all the unknown polynomials in (2), such as $\varphi_\ell(\mathbf{x}, \mathbf{c}_\ell), \sigma_l(\mathbf{x}), \lambda_{\ell\ell',i}(\mathbf{x})$. Note that since the coefficients of $\varphi_\ell(\mathbf{x}, \mathbf{c}_\ell), \eta_{\ell\ell'}(\mathbf{x})$ and $\nu_\ell(\mathbf{x})$ are unknown, some nonlinear terms that are products of these coefficients, will occur in the second and third constraints of (2), which yields a non-convex bilinear matrix inequalities (BMI) problem. In [15], to avoid this BMI problem we strengthened the second and third constraints in (1) to

$$g(\ell, \ell') \wedge \rho(\ell, \ell') \models \varphi_{\ell'}(\mathbf{x}', \mathbf{c}_{\ell'}) \geq 0 \quad \text{and} \quad \Psi(\ell) \models \dot{\varphi}_\ell(\mathbf{x}, \mathbf{c}_\ell) > 0,$$

respectively, which then results a linear matrix inequality (LMI) problem. For more details, please refer to Theorems 5 and 6 in [15]. In this paper, we will discuss in Section 4 how to handle the SOS programming (2) directly using the BMI solver or iterative method.

## 4. Approximate Solution from BMI Solver

In Section 3, we have reduced the problem of safety verification of a hybrid system to the SOS programming (2) involving BMI constraints.

Let us first show by an example on how to transform nonlinear parametric polynomial constraints into a BMI problem.

**Example 1.** Consider the system $\dot{x} = 2x$ with location invariant $\Psi = \{x \in \mathbb{R} : x^2 - 1 \leq 0\}$. From the discussion in Section 2, to find a polynomial $\varphi(x)$ satisfying $\varphi(x) \geq 0 \wedge \Psi \models \dot{\varphi}(x) \geq 0$, it suffices to find $\varphi(x)$ such that

$$\dot{\varphi}(x) = \phi_0(x) + \phi_1(x)(1 - x^2) + \phi_2(x)\varphi(x), \qquad (3)$$

where $\phi_0(x), \phi_1(x), \phi_2(x)$ are SOSes. Suppose that $\deg(\varphi) = 1, \deg(\phi_0) = 2$ and $\deg(\phi_1) = \deg(\phi_2) = 0$, and that $\varphi(x) = u_0 + u_1 x$, $\phi_1 = u_2$ and $\phi_2 = v_1$, with $u_0, u_1, u_2, v_1 \in \mathbb{R}$ parameters. From (3) we have

$$\phi_0(x) = u_2\, x^2 + (2\, u_1 - u_1\, v_1)x - u_2 - u_0\, v_1,$$

whose square matrix representation (SMR) [1] is $\phi_0(x) = Z^T Q Z$, where

$$Q = \begin{bmatrix} -u_2 - u_0\, v_1 & u_1 - \frac{1}{2} u_1\, v_1 \\ u_1 - \frac{1}{2} u_1\, v_1 & u_2 \end{bmatrix} \quad \text{and} \quad Z = \begin{bmatrix} 1 \\ x \end{bmatrix}.$$

Since all the $\phi_i(x)$ are SOSes, we have $u_2 \geq 0$, $v_1 \geq 0$ and $Q \succeq 0$, which can be expressed as one parametric positive semidefinite matrix

$$\mathcal{B}(u_0, u_1, u_2, v_1) = \begin{bmatrix} u_2 & 0 & 0 & 0 \\ 0 & v_1 & 0 & 0 \\ 0 & 0 & -u_2 - u_0\, v_1 & u_1 - \frac{1}{2} u_1\, v_1 \\ 0 & 0 & u_1 - \frac{1}{2} u_1\, v_1 & u_2 \end{bmatrix} \succeq 0.$$

Therefore, the constraint (3) is translated into a BMI constraint

$$\mathcal{B} = A_0 + v_1 A_1 + \sum_{i=0}^{2} u_i A_{2+i} + \sum_{j=0}^{2} u_j v_1 B_{j,1} \succeq 0,$$

where $A_i, B_{j,1}$ are constant symmetric matrices. □

Similar to Example 1, the SOS programming (2) can be transformed into a BMI problem of the form

$$\begin{aligned} \inf_{(\mathbf{u},\mathbf{v}) \in \mathbb{R}^{m+k}} & \quad F(\mathbf{u}, \mathbf{v}) \\ \text{s.t.} & \quad \mathcal{B}(\mathbf{u}, \mathbf{v}) = A_0 + \sum_{i=1}^{m} u_i A_i + \sum_{j=1}^{k} v_j A_{m+j} + \sum_{1 \leq i \leq m} \sum_{1 \leq j \leq k} u_i v_j B_{ij} \succeq 0. \end{aligned} \quad (4)$$

where $A_i, B_{ij}$ are constant symmetric matrices, $\mathbf{u} = (u_1, \ldots, u_m)$, $\mathbf{v} = (v_1, \ldots, v_k)$ are parameter coefficients of the SOSes occurring in the original SOS problem, and the objective function $F(\mathbf{u}, \mathbf{v})$ is a dummy objective function, which is commonly used for optimization problem with no objective functions.

Many methods can be used to solve the BMI problem (4) directly, such as interior-point constrained trust region method [14], an augmented Lagrangian strategy [9] and so on. A Matlab package PENBMI solver [10], which combines the (exterior) penalty and (interior) barrier method with the augmented Lagrangian method, can be applied directly on the BMI program (4). This can yield efficiently numerical solutions to the SOS programming (2).

Alternatively, observing in (4), $\mathcal{B}(\mathbf{u}, \mathbf{v})$ involves no crossing products like $u_i u_j$ and $v_i v_j$. Taking this special form into account, an iterative method can be applied by fixing $\mathbf{u}$ and $\mathbf{v}$ alternatively, which leads to a convex LMI problem [20, 26]. Below is an algorithm.

**Algorithm 1.** *BMI Solver based on Iteration*

1. [**Initialization**] Set $\mathbf{u} = \mathbf{u}_0$. Then (4) becomes to an LMI problem

$$\begin{aligned} \inf_{\mathbf{v} \in \mathbb{R}^k} & \quad F(\mathbf{u}_0, \mathbf{v}) \\ \text{s.t.} & \quad \mathcal{A}(\mathbf{u}_0, \mathbf{v}) = A_0 + \sum_{i=1}^{m} u_{i,0} A_i + \sum_{j=1}^{k} v_j A_{m+j} \succeq 0. \end{aligned}$$

   Suppose we obtain a feasible solution $\bar{\mathbf{v}}$ by solving the above LMI problem.

2. [**Fixing $\bar{\mathbf{v}}$**] Find an updated solution $\bar{\mathbf{u}}$ by solving the following LMI problem

$$\begin{aligned} \inf_{\mathbf{u} \in \mathbb{R}^m} & \quad F(\mathbf{u}, \bar{\mathbf{v}}) \\ \text{s.t.} & \quad \mathcal{A}(\mathbf{u}, \bar{\mathbf{v}}) = A_0 + \sum_{i=1}^{m} u_i A_i + \sum_{j=1}^{k} \bar{v}_j A_{m+j} \succeq 0. \end{aligned}$$

3. [**Fixing $\bar{\mathbf{u}}$**] Find an updated solution $\bar{\mathbf{v}}$ by solving the following LMI problem

$$\begin{aligned} \inf_{\mathbf{v} \in \mathbb{R}^k} & \quad F(\bar{\mathbf{u}}, \mathbf{v}) \\ \text{s.t.} & \quad \mathcal{A}(\bar{\mathbf{u}}, \mathbf{v}) = A_0 + \sum_{i=1}^{m} \bar{u}_i A_i + \sum_{j=1}^{k} v_j A_{m+j} \succeq 0. \end{aligned}$$

Repeat steps 2 and 3 until a feasible solution $(\bar{\mathbf{u}}, \bar{\mathbf{v}})$ is found such that $\mathcal{B}(\bar{\mathbf{u}}, \bar{\mathbf{v}}) \succeq 0$.

Remark that, although the convergence of Algorithm 1 can not be guaranteed, the iterative method is easier to implement than PENBMI solver. Moreover, from the experiments shown in Section 7, Algorithm 1 can yield a feasible solution $(\bar{\mathbf{u}}, \bar{\mathbf{v}})$ efficiently in practice.

# 5. Exact SOS Recovery

Since the SDP solvers in Matlab are running in fixed precision, the techniques in Section 4 only yield numerical solutions of (1). Due to round-off errors, $\varphi_\ell(\mathbf{x}, \mathbf{c}_\ell)$ may not be an invariant of the given hybrid system at location $\ell$, because the constraints in (1) may not hold exactly, as illustrated by the following example.

**Example 2.** [20, page 31] Consider the following nonlinear system

$$\left[\begin{array}{c} \dot{x}_1 \\ \dot{x}_2 \end{array}\right] = \left[\begin{array}{c} x_2 \\ -x_1 + \frac{1}{3}x_1^3 - x_2 \end{array}\right],$$

we want to verify that all trajectories of the system starting from the initial set $\Theta$ will never enter the unsafe region $X_u$, where

$$\Theta = \{(x_1, x_2) \in \mathbb{R}^2 : (x_1 - 1.5)^2 + x_2^2 \leq 0.25\}$$

and

$$X_u = \{(x_1, x_2) \in \mathbb{R}^2 : (x_1 + 1)^2 + (x_2 + 1)^2 \leq 0.16\}.$$

It suffices to find an invariant $\varphi(x_1, x_2)$ with rational coefficients, which satisfies all the conditions in (1). As stated in [15], we first set up an SDP system using LMI constraints. Apply the SDP solver to find a numerical polynomial invariant

$$\varphi(x_1, x_2) = -1.3686 + 0.62499\, x_1^2 + 1.0669\, x_1 x_2 + 1.5086\, x_2^2 - 0.56749\, x_1 x_2^2$$
$$- 0.15231\, x_2^3 - 0.10417\, x_1^4 - 0.35564\, x_1^3 x_2 - 0.23739\, x_1^2 x_2^2 - 0.24152\, x_1 x_2^3,$$

and some associated numerical positive semidefinite matrices $W^{[i]}$.

**Case 1:** If we convert the coefficients of $\varphi(x_1, x_2)$ to the corresponding rational coefficients separately, we obtain

$$\bar{\varphi}(x_1, x_2) = -\frac{6843}{5000} + \frac{62499}{100000}x_1^2 + \frac{10669}{10000}x_1 x_2 + \frac{7543}{5000}x_2^2 - \frac{56749}{100000}x_1 x_2^2$$
$$- \frac{15231}{100000}x_2^3 - \frac{10417}{100000}x_1^4 - \frac{8891}{25000}x_1^3 x_2 - \frac{23739}{100000}x_1^2 x_2^2 - \frac{3019}{12500}x_1 x_2^3.$$

However, $\bar{\varphi}(x_1, x_2)$ can not satisfy the conditions in (1) *exactly*, because there exists a sample point $p = (-\frac{127}{64}, -\frac{7}{8})$ such that the third constraint of (1) can not be satisfied. Therefore, $\bar{\varphi}(x_1, x_2)$ is not an *exact* invariant of this system.

**Case 2:** In our former papers [29, 15], we applied Gauss-Newton iteration and rational vector recovery to obtain solutions that satisfy the constraints in (1). This technique may fail in some cases. Let $\tau = 10^{-2}$ and the bound of the common denominator of the rational polynomial be 100. Then we obtain a rational polynomial

$$\tilde{\varphi}(x_1, x_2) = -\frac{15}{11} + \frac{5}{8}x_1^2 + \frac{47}{44}x_1 x_2 + \frac{133}{88}x_2^2 - \frac{25}{44}x_1 x_2^2$$
$$- \frac{13}{88}x_2^3 - \frac{9}{88}x_1^4 - \frac{31}{88}x_1^3 x_2 - \frac{21}{88}x_1^2 x_2^2 - \frac{21}{88}x_1 x_2^3.$$

However, the subsequent Gauss-Newton iteration and rational vector recovery techniques in [29, 15] failed in finding the associated positive semidefinite matrices that satisfy the constraints of the polynomial invariant *exactly*. The reason may lies in that we recover the coefficient vector of $\varphi(x_1, x_2)$ and the associated positive semidefinite matrices separately.

In the sequel, we will propose an improved algorithm to compute exact solutions of polynomial optimization problems with BMI constraints, through a modified Newton refinement and rational vector recovery technique applied on the coefficient vector $\mathbf{c}$ and the associated positive semidefinite matrices *simultaneously*.

## 5.1. Modified Newton Iteration

Suppose that applying PENBMI solver or Algorithm 1 in Section 4 yields numerical solutions that satisfy (1) approximately. Similar to [7], we now present the modified Newton refinement method to refine these solutions. Without loss of generality, we can reduce the problem (1) to the following problem

$$\begin{cases} \text{find} & \mathbf{c} \in \mathbb{R}^\nu \\ \text{s.t.} & \varphi_1(\mathbf{x}, \mathbf{c}) \geq 0, \\ & \varphi_3(\mathbf{x}, \mathbf{c}) \geq 0 \models \varphi_2(\mathbf{x}, \mathbf{c}) \geq 0, \end{cases} \quad (5)$$

where the coefficients of the polynomials $\varphi_i(\mathbf{x}, \mathbf{c}), 1 \leq i \leq 3$ are affine in $\mathbf{c}$. Then, based on SOS relaxation, the problem (5) can be further transformed into the following polynomial parametric optimization problem

$$\begin{cases} \text{find} & \mathbf{c} \in \mathbb{R}^\nu \\ \text{s.t.} & \varphi_1(\mathbf{x}, \mathbf{c}) = \mathbf{m}_1(\mathbf{x})^T \cdot W^{[1]} \cdot \mathbf{m}_1(\mathbf{x}), \\ & \varphi_2(\mathbf{x}, \mathbf{c}) = \mathbf{m}_2(\mathbf{x})^T \cdot W^{[2]} \cdot \mathbf{m}_2(\mathbf{x}) + (\mathbf{m}_3(\mathbf{x})^T \cdot W^{[3]} \cdot \mathbf{m}_3(\mathbf{x})) \cdot \varphi_3(\mathbf{x}, \mathbf{c}), \\ & W^{[i]} \succeq 0, 1 \leq i \leq 3, \end{cases} \quad (6)$$

involving both LMI and BMI constraints.

By solving the SDP system (6), we can obtain the numerical vector $\mathbf{c}$ and the approximate positive semidefinite matrices $W^{[i]}, 1 \leq i \leq 3$. We first convert $W^{[3]}$ to a nearby rational positive semidefinite matrix $\widetilde{W}^{[3]}$ by non-negative truncated $\mathrm{PLDL^T P^T}$-decomposition, in which all the diagonal entries of the corresponding diagonal matrix are preserved to be non-negative. Hereafter, denote by $\phi(\mathbf{x})$ the rational polynomial $\mathbf{m}_3(\mathbf{x})^T \cdot \widetilde{W}^{[3]} \cdot \mathbf{m}_3(\mathbf{x})$, and set $\theta = \|r_1(\mathbf{x})\|_2^2 + \|r_2(\mathbf{x})\|_2^2$, the backward error of the numerical solutions of (6), where

$$r_1(\mathbf{x}) = \varphi_1(\mathbf{x}, \mathbf{c}) - \mathbf{m}_1(\mathbf{x})^T \cdot W^{[1]} \cdot \mathbf{m}_1(\mathbf{x}),$$
$$r_2(\mathbf{x}) = \varphi_2(\mathbf{x}, \mathbf{c}) - \mathbf{m}_2(\mathbf{x})^T \cdot W^{[2]} \cdot \mathbf{m}_2(\mathbf{x}) - \phi(\mathbf{x})\varphi_3(\mathbf{x}, \mathbf{c}).$$

With the numerical $\mathbf{c}, W^{[1]}, W^{[2]}$ and the rational polynomial $\phi(\mathbf{x})$, we expand the two square matrix representations in (6) into their SOS forms respectively:

$$\varphi_1(\mathbf{x}, \mathbf{c}) \approx \sum_{i=1}^{t} \left( \sum_{\alpha} p_{i,\alpha} \mathbf{x}^\alpha \right)^2 \quad \text{and} \quad \varphi_2(\mathbf{x}, \mathbf{c}) \approx \sum_{j=1}^{k} \left( \sum_{\beta} q_{j,\beta} \mathbf{x}^\beta \right)^2 + \phi(\mathbf{x})\varphi_3(\mathbf{x}, \mathbf{c}), \quad (7)$$

where $t$ and $k$ are the ranks of $W^{[1]}$ and $W^{[2]}$ respectively. Apply Gauss-Newton iteration on two equations in (7) simultaneously to compute $\Delta \mathbf{c}, \Delta p_{i,\alpha}$ and $\Delta q_{j,\beta}$ such that

$$\varphi_1(\mathbf{x}, \mathbf{c} + \Delta \mathbf{c}) \approx \sum_{i=1}^{t} (\sum_\alpha (p_{i,\alpha} + \Delta p_{i,\alpha}) \mathbf{x}^\alpha)^2,$$
$$\varphi_2(\mathbf{x}, \mathbf{c} + \Delta \mathbf{c}) \approx \sum_{j=1}^{k} (\sum_\beta (q_{j,\beta} + \Delta q_{j,\beta}) \mathbf{x}^\beta)^2 + \phi(\mathbf{x})\varphi_3(\mathbf{x}, \mathbf{c} + \Delta \mathbf{c}).$$

We update the vector $\mathbf{c}$ and matrices $W^{[i]}$ by $\mathbf{c} + \Delta \mathbf{c}$ and $W^{[i]} + \Delta W^{[i]}, 1 \leq i \leq 2$, respectively, and terminate the Newton iteration when $\theta$ is less than the given tolerance $\tau$. In doing so, we will obtain the refined solution, the vector $\mathbf{c}$ and matrices $W^{[1]}, W^{[2]}$, of (6) such that

$$\left. \begin{array}{l} \varphi_1(\mathbf{x}, \mathbf{c}) - \mathbf{m}_1(\mathbf{x})^T \cdot W^{[1]} \cdot \mathbf{m}_1(\mathbf{x}) \approx 0, \\ \varphi_2(\mathbf{x}, \mathbf{c}) - \mathbf{m}_2(\mathbf{x})^T \cdot W^{[2]} \cdot \mathbf{m}_2(\mathbf{x}) - \varphi_3(\mathbf{x}, \mathbf{c})\phi(\mathbf{x}) \approx 0, \\ W = \begin{bmatrix} W^{[1]} & 0 \\ 0 & W^{[2]} \end{bmatrix} \succsim 0, \ W^T = W, \end{array} \right\} \quad (8)$$

and the backward error of the numerical solution satisfies $\theta < \tau$.

## 5.2. Exact Recovery

In this section, we will present two error-free algorithms to obtain a rational vector $\bar{\mathbf{c}}$, which satisfies the constraints in (5) exactly.

Discussed in Section 5.1, we can obtain a refined solution $\mathbf{c} \in \mathbb{R}^k$ by applying the modified Gauss-Newton iteration. Instead of recovering the entries of $\mathbf{c}$ separately, we can deploy simultaneous recovery technique [12] to achieve a rational vector $\bar{\mathbf{c}}$ near to $\mathbf{c}$ for a given bound $D$ of the common denominator of $\mathbf{c}$. Then we need verify whether $\bar{\mathbf{c}}$ is an exact solution of (5), that is, whether $\bar{\mathbf{c}}$ satisfies

$$\varphi_1(\mathbf{x}, \bar{\mathbf{c}}) \geq 0 \text{ and } \varphi_3(\mathbf{x}, \bar{\mathbf{c}}) \geq 0 \models \varphi_2(\mathbf{x}, \bar{\mathbf{c}}) \geq 0,$$

or equivalently, both

$$\varphi_1(\mathbf{x}, \bar{\mathbf{c}}) < 0 \text{ and } \varphi_3(\mathbf{x}, \bar{\mathbf{c}}) \geq 0 \wedge \varphi_2(\mathbf{x}, \bar{\mathbf{c}}) < 0$$

has no real solutions. Clearly, verifying the solution $\bar{\mathbf{c}}$ is equivalent to determining two constant semi-algebraic systems have no real solutions, which can be verified by Maple packages *RegularChains*, *DISCOVERER* [30] and *RAGLib* [3].

In this paper, we focus on retrieving exact SOS representations for (6). We will discuss how to recover from $\mathbf{c}, W^{[1]}, W^{[2]}$, the rational vector $\tilde{\mathbf{c}}$ and rational positive semidefinite matrices $\widetilde{W}^{[1]}$ and $\widetilde{W}^{[2]}$ that satisfy *exactly*

$$\varphi_1(\mathbf{x}, \tilde{\mathbf{c}}) - \mathbf{m}_1(\mathbf{x})^T \cdot \widetilde{W}^{[1]} \cdot \mathbf{m}_1(\mathbf{x}) = 0 \text{ and } \varphi_2(\mathbf{x}, \tilde{\mathbf{c}}) - \mathbf{m}_2(\mathbf{x})^T \cdot \widetilde{W}^{[2]} \cdot \mathbf{m}_2(\mathbf{x}) - \varphi_3(\mathbf{x}, \tilde{\mathbf{c}})\phi(\mathbf{x}) = 0. \quad (9)$$

Since the equations in (9) are affine in entries of $\tilde{\mathbf{c}}$ and $\widetilde{W}_1, \widetilde{W}_2$, one can define an affine linear hyperplane

$$\mathcal{L} = \Big\{ \mathbf{c}, W^{[1]}, W^{[2]} \mid \varphi_1(\mathbf{x}, \mathbf{c}) - \mathbf{m}_1(\mathbf{x})^T \cdot W^{[1]} \cdot \mathbf{m}_1(\mathbf{x}) = 0,$$
$$\varphi_2(\mathbf{x}, \mathbf{c}) - \mathbf{m}_2(\mathbf{x})^T \cdot W^{[2]} \cdot \mathbf{m}_2(\mathbf{x}) - \varphi_3(\mathbf{x}, \mathbf{c})\phi(\mathbf{x}) = 0 \Big\}. \quad (10)$$

Note that the hyperplane (10) can be constructed from a linear system $A\mathbf{y} = \mathbf{b}$, where $\mathbf{y}$ consists of the entries of $\mathbf{c}, W^{[1]}, W^{[2]}$. If $A$ has full row rank, such a hyperplane is guaranteed to exist. Then the rationalized SOS solutions of (6) can be computed by orthogonal projection if the matrix $W$ in (8) has full rank, or by the rational vector recovery otherwise.

*Case 1: W is of full rank*

Suppose the refined matrix $W$ in (8) is of full rank *numerically*, namely, the minimal eigenvalue of $W$ is greater than the given tolerance $\tau$. In this case, for a given bound $D$ of the common denominator, we apply orthogonal projection technique to obtain the rational vector $\tilde{\mathbf{c}}$ and rational matrix $\widetilde{W}$, which lie on the affine linear hyperplane defined by (10). This projection can be achieved by exactly solving the least squares problem:

$$\left. \begin{aligned} \min_{\tilde{\mathbf{c}}, \widetilde{W}} \; & \|\mathbf{c} - \tilde{\mathbf{c}}\|_2^2 + \|W - \widetilde{W}\|_F^2 \\ \text{s. t. } & \varphi_1(\mathbf{x}, \tilde{\mathbf{c}}) = \mathbf{m}_1(\mathbf{x})^T \cdot \widetilde{W}^{[1]} \cdot \mathbf{m}_1(\mathbf{x}), \\ & \varphi_2(\mathbf{x}, \tilde{\mathbf{c}}) = \mathbf{m}_2(\mathbf{x})^T \cdot \widetilde{W}^{[2]} \cdot \mathbf{m}_2(\mathbf{x}) + \phi(\mathbf{x})\varphi_3(\mathbf{x}, \tilde{\mathbf{c}}). \end{aligned} \right\} \quad (11)$$

For the rational solution $\tilde{\mathbf{c}}$ and $\widetilde{W}$ of (11), we compute the exact $\text{PLDL}^\text{T}\text{P}^\text{T}$-decomposition to check whether $\widetilde{W}$ is positive semidefinite. If so, then $\tilde{\mathbf{c}}$ is verified solution of (6).

**Theorem 3.** *Let $\mathbf{c}, W$ be the refined numerical solution of (6) with the backward error $\theta < \tau$, and $A\mathbf{y} = \mathbf{b}$ be the linear system associated to the hyperplane (10). Suppose that $\tilde{\mathbf{c}}$ and $\widetilde{W}$ are the optimal rational solutions of the least square problem (11). Let $\lambda \in \mathbb{R}_{>0}$ be the minimal eigenvalue of $W$. If $A$ has full row rank and $\lambda > 2\eta\kappa_2^2(A)\tau^2$ with $\eta = \|\mathbf{c}\|_2^2 + \|W\|_F^2$, then $\widetilde{W}$ is positive semidefinite, and $\tilde{\mathbf{c}}$ is a certified solution of (6).*

*Proof.* Clearly, $\tilde{\mathbf{c}}$ and $\widetilde{W}$ satisfy the equations in (9). Let $\mathbf{z}$ and $\tilde{\mathbf{z}}$ be the vectors consisting of the entries in $\mathbf{c}, W$ and those in $\tilde{\mathbf{c}}, \widetilde{W}$, respectively. Since $A$ has full row rank, we have $\|A\mathbf{z} - b\|_2^2 = \theta < \tau$, and then $A\tilde{\mathbf{z}} = \mathbf{b}$. According to the perturbation result in [4, Th.5.7.1] for full rank underdetermined systems, we have

$$\|\mathbf{z} - \tilde{\mathbf{z}}\| < (\kappa_2(A)\tau)\|\mathbf{z}\|_2 + O(\tau^2). \qquad (12)$$

From (12) and the assumption $\lambda > 2\,\eta\,\kappa_2^2(A)\,\tau^2$, we have $\|W - \widetilde{W}\|_F^2 \leq 2\,\|\mathbf{z} - \tilde{\mathbf{z}}\|_2^2 < \lambda$, where the last inequality follows from that $O(\tau^2)$ is negligible in comparison with $\eta$ when $\tau$ is very small. Let $\tilde{\lambda}$ be the minimal eigenvalue of $\widetilde{W}$. By Wielandt-Hoffman theorem [4, page 395], we have $|\tilde{\lambda} - \lambda| \leq \|W - \widetilde{W}\|_F^2 \leq \lambda$, which concludes that $\widetilde{W} \succeq 0$. □

**Remark 2.** For a numerical solution of (6), $2\eta\kappa_2^2(A)\tau^2$ in Theorem 3 is a constant. Therefore, if $A$ has full row rank and the given tolerance $\tau$ is small enough, then the orthogonal projection method by solving exactly least squares problem (11) will always find a rational positive semidefinite matrix $\widetilde{W}$, therefore yielding a certified rational solution $\tilde{\mathbf{c}}$ of (6).

*Case 2: W is singular*

It is pointed out in [8] that, the orthogonal projection by solving (11) may fail if the resulted rational matrix $\widetilde{W}$ satisfying (9) is positive semidefinite but not *strictly* positive definite. In this case, we need to explore rational vector recovering technique to obtain the rational $\tilde{\mathbf{c}}$ and $\widetilde{W}$. Unlike in [15], the technique to be presented will recover the rational vector $\tilde{\mathbf{c}}$ and matrix $\widetilde{W}$ *simultaneously*. More specifically, given a common denominator bound of the entries in $\tilde{\mathbf{c}}$ and $\widetilde{W}$, we employ the simultaneous Diophantine approximation algorithm [12] for $\mathbf{c}$ and the singular matrices in $\{W^{[1]}, W^{[2]}\}$. There are two cases to be addressed.

*Case 2.1:* Both $W^{[1]}$ and $W^{[2]}$ are singular. We apply rational vector recovery directly on $\mathbf{c}$ and $W$ simultaneously to obtain $\tilde{\mathbf{c}}$ and $\widetilde{W}$.

*Case 2.2:* Only one matrix, say, $W^{[1]}$ is singular. We apply vector recovery technique on $\mathbf{c}$ and $W^{[1]}$ simultaneously, and then check whether $\widetilde{W^{[1]}} \succeq 0$ by PLDL$^{\mathrm{T}}$P$^{\mathrm{T}}$-decomposition. If so, the rational matrix $\widetilde{W^{[2]}}$ can be obtained by orthogonal projection by solving the following exactly least squares problem:

$$\left. \begin{array}{l} \min_{\widetilde{W^{[2]}}} \|W^{[2]} - \widetilde{W^{[2]}}\|_F^2 \\ s.t. \quad \varphi_2(\mathbf{x}, \tilde{\mathbf{c}}) = \mathbf{m}_2(\mathbf{x})^T \cdot \widetilde{W^{[2]}} \cdot \mathbf{m}_2(\mathbf{x}) + \varphi_3(\mathbf{x}, \tilde{\mathbf{c}})\phi(\mathbf{x}). \end{array} \right\}$$

**Remark 3.** Similar to (6), we first apply the modified Newton iteration to compute the refined numerical solutions $\mathbf{c}$ and $W^{[i]}$ of (2) respectively. For the given tolerance and common denominator, we employ rational vector recovery technique simultaneously for $\mathbf{c}$ and all the singular matrices among the $W^{[i]}$, to find rational $\tilde{\mathbf{c}}$ and the associated rational positive semidefinite matrices. Finally, orthogonal projection is applied on the remaining nonsingular matrices to obtain the rational non-singular positive definite matrices.

### 5.3. Algorithm

The results in Sections 5.1 and 5.2 yield an algorithm to find exact solutions to (6).

**Algorithm 2.** *Verified Parametric Optimization Solver*

Input:
- ▶ a polynomial optimization problem of the form (6).
- ▶ $D \in \mathbb{Z}_{>0}$: the bound of the common denominator.
- ▶ $e \in \mathbb{Z}_{\geq 0}$: the degree bound $2e$ of the SOSes used to construct the SOS programming.
- ▶ $\tau \in \mathbb{R}_{>0}$: the given tolerance.

Output:
- ▶ the verified solution $\tilde{\mathbf{c}}$ of (6) with the $\widetilde{W^{[i]}}, 1 \leq i \leq 3$ positive semidefinite.

1. [**Compute the numerical solutions**] Set up the SOS programming (6) with the degree bound $2e$ and apply PENBMI solver or Algorithm 1 to compute its numerical solutions. If the system has no feasible solutions, return "we can't find solutions of (6) with the given degree bound $2e$". Otherwise, obtain **c** and $W^{[i]} \gtrapprox 0, 1 \leq i \leq 3$.

2. [**Compute the verified solution $\tilde{\mathbf{c}}$**]

   (2.1) Convert $W^{[3]}$ to a nearby rational positive semidefinite matrix $\widetilde{W}^{[3]}$ by non-negative truncated $\mathrm{PLDL^T P^T}$-decomposition.

   (2.2) For the tolerance $\tau$, apply the modified Newton iteration to refine **c** and $W^{[1]}, W^{[2]}$.

   (2.3) Determine the singularity of $W^{[1]}$ and $W^{[2]}$ with respect to $\tau$.

   Case 1: Both $W^{[1]}$ and $W^{[2]}$ are of full rank. The rational vector $\tilde{\mathbf{c}}$, and rational matrices $\widetilde{W}^{[1]}$ and $\widetilde{W}^{[2]}$ can be obtained by orthogonal projection for a given bound $D$ of the common denominator Check if $\widetilde{W}^{[i]} \succeq 0$ for $i = 1, 2$. If so, return $\tilde{\mathbf{c}}$, and $\widetilde{W}^{[i]}, 1 \leq i \leq 3$. Otherwise, return "we can't find the solution of (6) with the given degree bound".

   Case 2: Both $W^{[1]}$ and $W^{[2]}$ are singular. Obtain $\tilde{\mathbf{c}}$ and $\widetilde{W}^{[i]}$ as described in *Case 2.1* of Section 5.2.

   Case 3: Only one of $W^{[1]}$ and $W^{[2]}$ is singular. Obtain $\tilde{\mathbf{c}}$ and $\widetilde{W}^{[i]}$ as described in *Case 2.2* of Section 5.2.

## 6. Computational Issue

In practice, it may happen that the unsafe region is too big to find an appropriate invariant for verifying the safety of the given system. To deal with this issue, we can divide the original unsafe region into two parts, which can be achieved by bisection method through computing the minimum and maximum values of some variable $x_i$. Then we compute the invariants that verify safety for two unsafe sub-regions $X_{u,1}, X_{u,2}$, respectively. This procedure can be easily repeated until the safety property have been verified for all of those unsafe sub-regions. The process for splitting the initial set $\Theta$ can be handled similarly.

The following example is presented to illustrate the above technique.

**Example 3.** [22, Example 2] Consider the following nonlinear system

$$\begin{bmatrix} \dot{x}_1 \\ \dot{x}_2 \end{bmatrix} = \begin{bmatrix} x_1 - x_2 \\ x_1 + x_2 \end{bmatrix},$$

within the region $\Psi = \{(x_1, x_2) \in \mathbb{R}^2 : 0 \leq x_1 \leq 4 \wedge 0 \leq x_2 \leq 4\}$. We want to verify that all trajectories of the system starting from the initial set $\Theta$ will never enter the unsafe region $X_u$, where $\Theta = \{(x_1, x_2) \in \mathbb{R}^2 : 2.5 \leq x_1 \leq 3 \wedge x_2 = 0\}$ and $X_u = \{(x_1, x_2) \in \mathbb{R}^2 : x_1 \leq 2\}$.

For $X_u$, PENBMI solver or Algorithm 1 can not yield a feasible solution of the associated SOS program. Therefore, it is impossible to obtain the invariant to verify the safety property for this system. Here we split the original unsafe region into $X_{u,1} = \{(x_1, x_2) \in \mathbb{R}^2 : x_1 \leq 2 \wedge x_2 \leq 0\}$ and $X_{u,2} = \{(x_1, x_2) \in \mathbb{R}^2 : x_1 \leq 2 \wedge -x_2 \leq 0\}$. Then, our certified method, combining modified Gauss-Newton refinement with exact recovery technique, can yield the verified invariants with degree 2 for $X_{u,1}$ and $X_{u,2}$, respectively. Thus the safety of this system is guaranteed.

## 7. Experiments

Let us present some examples of safety verification for hybrid systems.

**Example 4 (Example 2 Revisited).** Consider the safety verification problem of Example 2. We now apply Algorithm 2 to compute two invariants $\tilde{\varphi}_1$ and $\tilde{\varphi}_2$, that are subject to the strengthened LMI constraints in [15] and the BMI constraints in (2), respectively.

According to Theorem 6 in [15], we construct an SOS system involving only LMI constraints. Applying the improved Gauss-Newton iteration and rational vector recovery techniques, we find an exact invariant $\tilde{\varphi}_1(x_1, x_2) \geq 0$ until $\deg(\tilde{\varphi}_1) = 4$, and the corresponding SOSes. Here we only list the invariant

$$\tilde{\varphi}_1(x_1, x_2) = -\frac{53}{39} + \frac{8}{13} x_1^2 + \frac{59}{39} x_2^2 - \frac{2}{13} x_2^3 - \frac{4}{39} x_1^4$$
$$+ \frac{14}{13} x_1 x_2 - \frac{22}{39} x_1 x_2^2 - \frac{14}{39} x_1^3 x_2 - \frac{3}{13} x_1^2 x_2^2 - \frac{3}{13} x_1 x_2^3.$$

The invariant $\tilde{\varphi}_1(x_1, x_2) \geq 0$ guarantees the safety of the given system.

Alternatively, from Theorem 1 we construct another SOS system with BMI constraints and then apply Algorithm 2 to obtain an exact invariant with degree 2:

$$\widetilde{\varphi}_2(x_1, x_2) = -\frac{151}{99} - \frac{62}{33} x_2 - \frac{152}{99} x_1 - \frac{106}{99} x_1 x_2 - \frac{4}{9} x_1^2.$$

This proves the safety of the given system.

Consider again the system in Example 2 with the same initial set $\Theta$ but a larger unsafe region $X_u = \{(x_1, x_2) \in \mathbb{R}^2 : (x_1 + 1)^2 + (x_2 + 1)^2 \leq 1\}$. Similarly, we construct the SOS systems with LMI constraint and BMI constraint respectively, and apply Algorithm 2 to compute the invariants $\widetilde{\varphi}_1(x_1, x_2) \geq 0$ with $\deg(\widetilde{\varphi}_1) = 6$, and $\widetilde{\varphi}_2(x_1, x_2) \geq 0$ with $\deg(\widetilde{\varphi}_2) = 4$, where

$$\widetilde{\varphi}_1(x_1, x_2) = \frac{1714}{3209} + \frac{1856}{3209} x_2^3 + \frac{6160}{3209} x_1^3 - \frac{3927}{3209} x_2^2 - \frac{1507}{3209} x_1 x_2^3 + \cdots - \frac{75}{3209} x_2^5 + \frac{426}{3209} x_1^6,$$

$$\widetilde{\varphi}_2(x_1, x_2) = -\frac{16}{25} - \frac{73}{50} x_1 + \frac{71}{50} x_2 - \frac{147}{100} x_1^2 + \cdots - \frac{299}{100} x_1^2 x_2^2 - \frac{347}{100} x_1 x_2^3 - \frac{211}{100} x_1^4. \quad \square$$

From Example 4, we see that the invariants obtained from the BMI constraints are of lower degree than those obtained from the LMI constraints.

**Example 5.** Consider a hybrid system [20] depicted in Figure 1, where

$$f_1(\mathbf{x}, d) = \begin{bmatrix} x_2 \\ -x_1 + x_3 \\ x_1 + (2x_2 + 3x_3)(1 + x_3^2) + d \end{bmatrix}, \quad f_2(\mathbf{x}, d) = \begin{bmatrix} x_2 \\ -x_1 + x_3 \\ -x_1 - 2x_2 - 3x_3 + d \end{bmatrix},$$

with parameter $d$ satisfies $-1 \leq d \leq 1$. The system starts in location $\ell_1$, with an initial state in

$$\Theta = \{(x_1, x_2, x_3) \in \mathbb{R}^3 : x_1^2 + x_2^2 + x_3^2 \leq 0.01\}.$$

Our task is to verify the system never reach the states of

$$X_u(\ell_2) = \{(x_1, x_2, x_3) \in \mathbb{R}^3 : 5 \leq x_1 \leq 5.1 \vee -5.1 \leq x_1 \leq -5\}.$$

To prove the safety of the hybrid system, it suffices to find the corresponding invariant polynomials $\varphi_1(\mathbf{x})$ and $\varphi_2(\mathbf{x})$ at locations $\ell_1$ and $\ell_2$, which satisfy all the conditions in Theorem 1.

Applying Algorithm 2, we obtain the invariant polynomials with rational coefficients

$$\widetilde{\varphi}_1(\mathbf{x}) = -\frac{53}{44} - \frac{39}{88} x_1 x_2 + \frac{5}{88} x_1 x_3 - \frac{1}{88} x_2 x_3 - \frac{1}{44} x_1^2 - \frac{3}{88} x_2^2 - \frac{3}{88} x_3^2,$$

$$\widetilde{\varphi}_2(\mathbf{x}) = -\frac{129}{22} - \frac{1}{88} x_1 x_2 + \frac{1}{88} x_1 x_3 + \frac{1}{88} x_2 x_3 + \frac{1}{88} x_1^2 + \frac{1}{88} x_2^2 + \frac{1}{88} x_3^2.$$

Moreover, $\widetilde{\varphi}_1(\mathbf{x})$ and $\widetilde{\varphi}_1(\mathbf{x})$ satisfy the conditions in Theorem 1 exactly. Therefore, the invariants can guarantee the safety of the hybrid system.

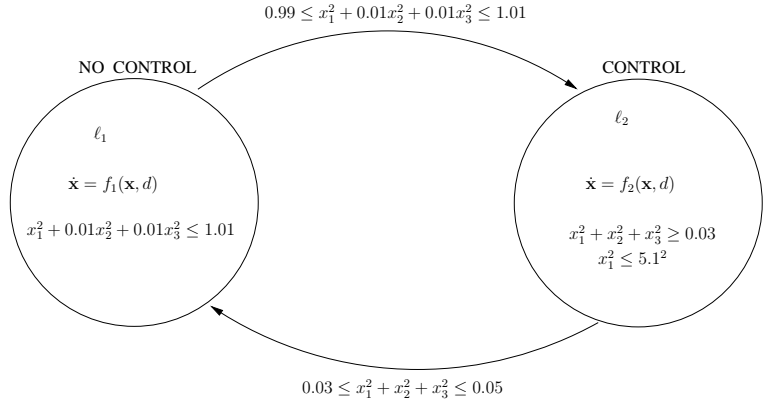

Figure 1: Hybrid system of Example 3

Table 1 shows the performance of Algorithm 2, for safety verification of some interesting benchmark examples. All the computations have been performed on an Intel Core 2 Duo 2.0 GHz processor with 2GB of memory. Examples 1–3 correspond to Examples CLOCK, FOCUS, ECO in [23], and Examples 4–7 correspond to [21, Example 3], [32, Example 3], [22, Example 11] and [31, Example 2]. All these examples except Example 2 are nonlinear systems. In all the associated SOS programmings, the degree bound of SOSes is $e = 6$ and we set $D = 1000$ and $\tau = 10^{-10}$ in Algorithm 2. Here $|\ell|$ and $n$ denote the number of locations and the number of system variables respectively; *BMI Solver* refers to the method used to obtain the numerical solutions of the given BMI problems; *Num. Para.* is the number of decision variables appearing in the BMI problem; $\deg(\widetilde{\varphi})$ and $\deg(\widetilde{\varphi}_{\text{LMI}})$ denote the degrees of invariants computed by Algorithm 2 and by the algorithm used in [15] with LMI constraints, respectively. *Fail* means that the algorithm in [15] fails to find invariants with degree $\leq 6$. *Time* is that for the entire computation run Algorithm 2 in seconds.

| Ex. | $|\ell|$ | $n$ | BMI solver | Num. Para. | $\deg(\widetilde{\varphi})$ | $\deg(\widetilde{\varphi}_{\text{LMI}})$ | Time (s) |
|---|---|---|---|---|---|---|---|
| 1 | 1 | 2 | PENBMI | 30 | 2 | 3 | 6.17 |
| 2 | 1 | 2 | Alg. 1 | 16 | 2 | Fail | 4.56 |
| 3 | 2 | 2 | Alg. 1 | 70 | 2 | 2 | 14.21 |
| 4 | 1 | 2 | PENBMI | 18 | 2 | Fail | 4.31 |
| 5 | 1 | 2 | PENBMI | 41 | 3 | Fail | 7.99 |
| 6 | 1 | 3 | Alg. 1 | 21 | 2 | 2 | 5.45 |
| 7 | 1 | 2 | Alg. 1 | 24 | 2 | 2 | 5.62 |

Table 1: Algorithm Performance on Benchmarks

## 8. Conclusion

In this paper, we present a symbolic-numeric method on safety verification of hybrid systems. A numerical invariant of a hybrid system can be obtained by solving a bilinear SOS programming via PENBMI solver or iterative method. Then a method based on modified Newton iteration and rational vector recovery techniques is deployed to obtain exact polynomial invariantswith rational number coefficients. Some experimental results are given to show the efficiency of our method.

## Acknowledgments

We would like to acknowledge helpful discussions with Professor Lu Yang and Professor Wensheng Yu.